\documentclass[11pt,a4paper]{article}

\usepackage{amssymb}
\usepackage{amsmath}
\usepackage[pdftex]{graphicx}
\usepackage{graphics}
\usepackage{epsfig}
\usepackage{color}

\newcommand{\be}{\begin{equation}}
\newcommand{\ee}{\end{equation}}
\numberwithin{equation}{section}

\newtheorem{theor}{Theorem}[section] 
\newtheorem{coro}{Corollary}

\title{\bf The Birman-Schwinger operator for a parabolic quantum well in a zero-thickness layer in the presence of a two-dimensional attractive Gaussian impurity}

\author{
S. Albeverio$^{1,2}$\footnote{albeverio@iam.uni-bonn.de},
S. Fassari$^{1,3}$\footnote{silvestro.fassari@uva.es}, 
M. Gadella$^4$\footnote{manuelgadella1@gmail.com},  
L.M. Nieto$^4$\footnote{luismiguel.nieto.calzada@uva.es}, and 
F. Rinaldi$^{1,3}$\footnote{f.rinaldi@unimarconi.it}
\\ [2ex]
\footnotesize \sl $^1$CERFIM, PO Box 1132, CH-6601 Locarno, Switzerland.\\
\footnotesize \sl $^2$Institut f\"ur Angewandte Mathematik, HCM, Universit\"at Bonn, \\  
\footnotesize \sl Endenicheralee 60, D-53115 Bonn, Germany\\
\footnotesize \sl $^3$Dipartimento di Fisica Nucleare, Subnucleare e delle Radiazioni, \\ 
\footnotesize \sl Univ. degli Studi Guglielmo Marconi,Via Plinio 44, I-00193 Rome, Italy.\\
\footnotesize \sl $^4$Departamento de F\'{\i}sica Te\'{o}rica, At\'{o}mica y \'{O}ptica, and IMUVA,\\ 
\footnotesize \sl U. de Valladolid, 47011 Valladolid, Spain.   
}

\begin{document}

\maketitle

\begin{abstract}
 In this note we consider a quantum mechanical particle moving inside an infinitesimally thin layer constrained by a parabolic well in the  $x$-direction and, moreover, in the presence of an impurity modelled by an attractive Gaussian potential. We investigate the Birman-Schwinger operator associated to a model assuming the presence of a Gaussian impurity inside the layer and prove that such an integral operator is Hilbert-Schmidt, which allows the use of the modified Fredholm determinant in order to compute the bound states created by the impurity. Furthermore, we consider the case where the Gaussian potential degenerates to a $\delta$-potential in  the $x$-direction and a Gaussian potential in the $y$-direction. We construct the corresponding self-adjoint Hamiltonian and prove that it is the limit in the norm resolvent sense of a sequence of corresponding Hamiltonians with suitably scaled Gaussian potentials. Satisfactory bounds on the ground state energies of all Hamiltonians involved are exhibited.
\end{abstract}


\newpage

\section{Introduction}

The study of point potentials in Quantum Physics has recently received a lot of attention for a wide range of interests. First of all, point potentials serve as solvable or quasi-solvable models that approximate the action of intense and very short range potentials \cite{AGHH,SEB,BHS}. They have been used to model several kinds of extra thin structures \cite{ZOL,ZOL1}, to mimic point defects in materials, or to study heterostructures \cite{GHNN,KP,KP1,KP2}. In addition, point potentials play a role in modelling impurities in quantum field theory \cite{MMM,AGM,MKB,MM}. Furthermore, they play an important role after a recent interpretation of the Casimir effect \cite{BOR,AM}. The unexpected relations between contact potentials and group theory should also be noted \cite{GMMN}. They also play a role in modelling Kronig-Penney crystals in condensed matter physics in various dimensions \cite{AGHH,LAR,F,F1,F2,F3,FK}. 

More examples of physical applications of this kind of interactions are: Bose-Einstein condensation in a harmonic trap with a tight and deep "dimple'' potential, modelled by a Dirac delta function \cite{HAY}; non-perturbative study of the entanglement of two directed polymers subjected to repulsive interactions given by a Dirac $\delta$-function potential \cite{FRV}; a periodic array of Dirac delta interactions is useful to investigate the light propagation in a one-dimensional realistic dielectric superlattice, which has been investigated for the transverse electric and magnetic fields and for omnidirectional polarisation modes \cite{AHS,ZS,MR}.  

One-dimensional quantum models with contact interactions are also useful to study a wide range of quantum properties, including scattering, since these models are quite often solvable. They also serve to acquire experience in order to analyse systems with contact potentials in higher dimensions. However, an important difference is to be pointed out:  while one-dimensional contact potentials are usually defined through matching conditions at isolated points, their proper definition  in higher dimensions requires a process of regularisation.  

Quantum two-dimensional systems are particularly interesting for their physical applications. In this particular context, the graphene deserves a special mention because of its importance, although this is not the only one two-dimensional quantum system of interest in physics. From a theoretical point of view, quantum theory in two dimensions has not yet been developed to the same extent of its one-dimensional and three-dimensional analogues, in spite of its enormous interest. Although two-dimensional quantum systems look rather simple, due to the presence of logarithmic singularities in the resolvent kernel of their free Hamiltonian, their level of complexity and, hence, their difficulty of study is far higher than the one-dimensional case (and, to a great extent, the three-dimensional case). 

An important contribution to the understanding of two-dimensional quantum mechanics was provided by Duclos' article \cite{DP} on the two-dimensional hydrogen atom perturbed by a point interaction, a model that had not been dealt with in \cite{AGHH}. One of the main results of that paper is that the free Hamiltonian of that model, namely that of the 2D  hydrogen atom, was rigorously shown to be the norm resolvent limit of the Hamiltonian of the 3D Hamiltonian of the  hydrogen atom confined to an infinite planar slab of width $a>0$ as $a\rightarrow 0_{+}$. 

A remarkable feature of two-dimensional models with contact interactions, manifesting itself even in the simple case of the negative Laplacian perturbed by a point interaction, is represented by their peculiar dependence of the bound state energies on the coupling constant. As is well known, the one-dimensional model exhibits a single bound state only when the point interaction is attractive and the eigenvalue is a quadratic function of the strength of the interaction. The three-dimensional case also exhibits a single bound state only in the attractive case but the eigenvalue depends quadratically on the reciprocal of the renormalised coupling constant. In two dimensions the bound state keeps existing even if the contact interaction is repulsive and the dependence becomes exponential (see \cite{AGHH}). The latter behaviour is confirmed even when a confinement potential is present in addition to the contact potential, which physically mimics the presence of impurities or thin barriers in the material inside which the quantum particle is moving \cite{FGGN2,FGGNR}.

In this note, partly motivated by Duclos' paper, we intend to study a different two-dimensional model with the free Hamiltonian given by: 

\be\label{H_0}
H_0=\left(-\frac 1{2} \frac {d^2}{dx^2}+\frac{x^2}2\right)-\frac 1{2}\frac {d^2}{dy^2}\,,
\ee
to which we add an attractive impurity assumed to be modelled by the isotropic Gaussian potential 
\be\label{W}
W(x,y)=-\lambda V(x,y)=-\lambda e^{-(x^2+y^2)},\quad \lambda>0\,,
\ee 
so that the total Hamiltonian is 
\begin{equation}\label{1111} 
H_\lambda=H_0+W(x,y)= H_0-\lambda V(x,y)=H_0-\lambda e^{-(x^2+y^2)},\quad \lambda>0.
\end{equation} 

It is worth mentioning that the recent literature \cite{MFRM,FGNR,FER,NAN} has shown a renewed interest in the spectral analysis of the one-dimensional Hamiltonian with a Gaussian potential, namely
\begin{equation}\label{1}
H:=-\frac12 \frac{d^2}{dx^2} -\lambda\,e^{-x^2/2},\quad \lambda>0.
\end{equation}
Therefore, \eqref{1111} could also be regarded as a possible two-dimensional generalisation of  \eqref{1}.

At this point, it is interesting to recall that a three-dimensional material with confinement in only one dimension is said to be a quantum well \cite{Har}, while a 3D material with two-dimensional confinement is called a quantum wire. Therefore, in the limiting case of a quantum well inside a layer with zero thickness, it makes sense to consider the model in which the confining potential is parabolic. Due to the mathematical subtleties required, in this note we have chosen to omit the proof of the resolvent convergence of the Hamiltonian of a three-dimensional parabolic quantum well inside a thin layer to the 2D Hamiltonian \eqref{1111} as the thickness of the layer vanishes. 
 
Instead we start directly by writing the Green function of the two-dimensional Hamiltonian with a one-dimensional harmonic potential. Once a perturbation given by an attractive two-dimensional Gaussian potential is added, we study the properties of the corresponding Birman-Schwinger operator, that is to say the crucial part in the interaction term of the resolvent of the perturbed Hamiltonian. We remind the reader that the resolvent is the key to obtain the energy eigenvalues. We later consider the model in which the 2D Gaussian impurity potential gets replaced by one having a Dirac delta for the coordinate subjected to the harmonic confinement maintaining the Gaussian character for the other coordinate. 

It may be worth pointing out that the Hamiltonian  \eqref{H_0} has been used by Dell'Antonio and collaborators \cite{GF} as the free Hamiltonian in the model of a quantum system consisting of two one-dimensional particles, one of which is harmonically bound to its equilibrium position, mutually interacting by means of the contact interaction $\delta (x-y)$. In other words, the interaction studied in \cite{GF} will be replaced by the one in \eqref{W}.

Solving the eigenvalue problem for this kind of Hamiltonians is not, in general, an easy task and often requires rather sophisticated tools. One of the most widely used is the Birman-Schwinger operator, namely the integral operator 
\begin{equation}\label{birmansch}
 B_E=(\text{sgn}\, W) |W|^\frac1{2}(H_0-E)^{-1}|W|^\frac 1{2}, 
 \end{equation} 

\noindent and the related technique: as in most applications $ B_E$ can be shown to be compact, the solutions of the eigenvalue problem for the Hamiltonian are given by those values of $ E$ for which  $ B_E$ has an eigenvalue equal to -1 (see \cite{KLA,KLA1,FK} and references therein as well as~\cite{RSVI}, p. 99). Therefore, the detailed study of the properties of the Birman-Schwinger operator arising from our model is quite relevant. In the present note, we show that the Birman-Schwinger operator is Hilbert-Schmidt. We also show that $ H_\lambda$ is self-adjoint and bounded from below. 

In addition, $H_\lambda$ has a special relation with a kind of two-dimensional contact operator that will be studied in Section 2.1. This is given by the Hamiltonian described heuristically by
\begin{equation}\label{4}
H^\delta_\lambda=H_0-\lambda\sqrt\pi\,\delta(x)\,e^{-y^2},\lambda>0 ,
\end{equation}
where $\delta(x)$ is the Dirac delta centred at the origin. We show that $H^\delta_\lambda$ is self-adjoint on a natural domain and can be obtained as the limit in the norm resolvent sense as $n\longmapsto\infty$ of the following sequence of Hamiltonians:
\begin{equation}\label{5}
H_{n,\lambda}:=H_0-\lambda n\,e^{-(n^2x^2+y^2)}, \lambda>0 ,
\end{equation}
thus with Gaussian type potentials (as was the case for $H_\lambda$) which become increasingly more attractive and anisotropic as $n$ goes to infinity.

Finally, it will be shown that the Hamiltonian $ H_\lambda$ (resp. $H^\delta_\lambda$) is bounded from below and its lower bound can be obtained using a certain transcendental equation.

\section{The Birman-Schwinger operator for our model}

Starting from the Hamiltonian $ H_0$ in \eqref{H_0}, it is rather straightforward to infer that the associated Green function, namely the integral kernel of the resolvent operator, reads for any $E<\frac 1{2}$:
\begin{equation}\label{0}
(H_0-E)^{-1}(x,x',y,y')=\sum _{n=0}^{\infty }\frac {e^{- \sqrt{2 \left(n+\frac 1{2}-E\right)} \, \left|y-y'\right|}}{\sqrt{2 \left(n+\frac 1{2}-E\right)}}\, \phi _{n}(x)\phi _{n}(x'),
\end{equation} 
where $ \phi _{n}(x)$ is the normalised $n$-th eigenfunction of the one-dimensional harmonic oscillator.

Once the above attractive Gaussian perturbation \eqref{W} is added, the total Hamiltonian is $H_\lambda$ in \eqref{1111}.
Therefore, its associated Birman-Schwinger integral kernel  \cite {MFRM,FGNR,KLA,KLA1,FK,F 95,FR 09} given by \eqref{birmansch} is:
\begin{eqnarray}\label{2}
 B_E\!\!&\!\!=\!\!&\!\! 
 -\lambda\, \tilde{B}_E \left(x,x_1,y,y_1\right)=
-\lambda \left|V\right|^\frac1{2}(H_0-E)^{-1}\left|V\right|^\frac 1{2}(x,x_1,y,y_1)=\nonumber\\ [1ex]
\!\!&\!\!=\!\!&\!\! -\lambda e^{-(x^2+y^2)/2}\left[\sum_{n=0}^{\infty }\frac {e^{-\sqrt{2 \left(n+\frac 1{2}-E\right)} \left|y-y_1 \right|}}{\sqrt{2 \left(n+\frac 1{2}-E\right)}} \phi_{n}(x)\phi_{n}(x_1)\right] e^{-(x_1^2+y_1^2)/2},
\end{eqnarray}

The main goal of this brief note is to rigorously prove that such an integral operator is Hilbert-Schmidt, that is to say $\text{tr}(\tilde{B}_E^2)<\infty$, given the evident positivity of the operator $ \tilde{B}_E$ (and our choice $\lambda>0$). As the kernel of the positive operator $ \tilde{B}_E^2$ is clearly
\begin{eqnarray}\label{3}
 \tilde{B}_E^2 \left(x,x_2,y,y_2\right) \!\!&\!\!=\!\!&\!\!  e^{-(x^2+y^2)/2} \left\{ \int_{-\infty}^{\infty} \int_{-\infty}^{\infty} \left[\sum _{m=0}^{\infty }\frac {e^{- \sqrt{2 \left(m+\frac 1{2}-E\right)} \left|y-y' \right|}}{\sqrt{2 \left(m+\frac 1{2}-E\right)}} \phi _{m}(x)\phi _{m}(x')\right] \right. \\
\!\!&\!\!\!\!&\!\!  \times \left.   e^{-(x'^2+y'^2)}\left[\sum _{n=0}^{\infty }\frac {e^{- \sqrt{2 \left(n+\frac 1{2}-E\right)} \left|y'-y_2 \right|}}{\sqrt{2 \left(n+\frac 1{2}-E\right)}} \phi _{n}(x')\phi _{n}(x_2)\right] dx'dy' \right\} e^{-(x_2^2+y_2^2)/2},
\nonumber
\end{eqnarray}
its trace reads
\begin{eqnarray}
\text{tr}( \tilde{B}_E^2) \!\!&\!\!= \!\!&\!\!  \int_{-\infty}^{\infty}\int_{-\infty}^{\infty} \tilde{B}_E^2 \left(x,x,y,y\right)dxdy=
\int_{-\infty}^{\infty}\int_{-\infty}^{\infty}\int_{-\infty}^{\infty} \int_{-\infty}^{\infty}  dxdydx'dy'\, e^{-(x^2+y^2)} e^{-(x'^2+y'^2)}  \nonumber  \\
&& \qquad\times \sum _{m=0}^{\infty }\sum _{n=0}^{\infty }\frac {e^{- \left(\sqrt{2 (m+\frac 1{2}-E)}+ \sqrt{2 (n+\frac 1{2}-E)}\right)\left|y-y' \right|}}
{2\sqrt{(m+\frac 1{2}-E)(n+\frac 1{2}-E)}}\phi _{m}(x)\phi _{m}(x') \phi _{n}(x)\phi _{n}(x').
\label{4}
\end{eqnarray}
The latter multiple integral can be rewritten as:
\begin{equation}\label{5}
\sum _{m=0}^{\infty }\sum _{n=0}^{\infty } \left[ \int_{-\infty}^{\infty}\int_{-\infty}^{\infty}   e^{-y^2}\frac {e^{- \left(\sqrt{2 (m+\frac 1{2}-E)}+ \sqrt{2 (n+\frac 1{2}-E)}\right)\left|y-y' \right|}}{2\sqrt{(m+\frac 1{2}-E)(n+\frac 1{2}-E)}}e^{-y'^2}dydy'\right]\left\langle\phi_{m}, e^{-(\cdot)^2}\phi_{n}\right\rangle^2 
\end{equation}
where $\langle f,g\rangle$ denotes the standard scalar product of the two functions.

Let us consider the double integral in \eqref{5}. With the notation,
\begin{equation}
f(y-y'):= e^{- \left(\sqrt{2 (m+\frac 1{2}-E)}+ \sqrt{2 (n+\frac 1{2}-E)}\right)\left|y-y' \right|}\,,\qquad g(y')=e^{-{y'}^2}\,,
\end{equation}
the second integral becomes the convolution $(f*g)(y)$, so that the double integral may be written as
\begin{eqnarray}\label{2m}
\frac 1{2\sqrt{(m+\frac 1{2}-E)(n+\frac 1{2}-E)}} \int_{-\infty}^\infty e^{-y^2}\, [(f*g)(y)]\,dy\,.
\end{eqnarray}
Using the Schwarz inequality \eqref{2m} is smaller than or equal to
\begin{equation}\label{3m}
\frac 1{2\sqrt{(m+\frac 1{2}-E)(n+\frac 1{2}-E)}} ||e^{-(\cdot)^2}||_2\, || f*g||_2\,,
\end{equation}
where $||\cdot||_p$ denotes the norm in $L^p(\mathbb R)$. Young's inequality \cite{RS2} shows that 
\begin{equation}\label{4m}
|| f*g||_r \le||f||_p\,||g||_q\,, \qquad {\rm with} \qquad \frac1p+\frac1q=\frac1r+1\,.
\end{equation}
Therefore, with $p=r=2$ and $q=1$, it follows that \eqref{3m} is smaller than or equal to
\begin{equation}\label{44m}
\frac 1{2\sqrt{(m+\frac 1{2}-E)(n+\frac 1{2}-E)}}\, ||e^{-(\cdot)^2}||^2_2\, ||f||_1\,.
\end{equation}
The two norms in \eqref{4m} yield two integrals which can be easily computed, so as to obtain
\begin{eqnarray}\label{5m}
\frac {\sqrt{\pi}}{2^{\frac 3{2}}\sqrt{(m+\frac 1{2}-E)(n+\frac 1{2}-E)}\left(\sqrt{2 (m+\frac 1{2}-E)}+ \sqrt{2 (n+\frac 1{2}-E)}\right)} 
\nonumber\\[2ex] 
\le  \frac {\sqrt{\pi}}{4(m+\frac 1{2}-E)^\frac 3{4}(n+\frac 1{2}-E)^\frac 3{4}}\,.
\end{eqnarray}

Hence, the trace \eqref{4} is bounded by:
\begin{equation}\label{7}
\text{tr}( \tilde{B}_E^2)\leq \frac {\sqrt{\pi}}{4}\sum _{m=0}^{\infty }\sum _{n=0}^{\infty }\frac {\left\langle \phi_{m}, e^{-(\cdot)^2}\phi_{n} \right\rangle^2}{(m+\frac 1{2}-E)^\frac 3{4}(n+\frac 1{2}-E)^\frac 3{4}}=\frac {\pi^\frac 3{2}}{4}\sum _{m=0}^{\infty }\sum _{n=0}^{\infty }\frac {\left\langle \phi_{m}\phi_0, \phi_0 \phi_{n} \right\rangle^2}{(m+\frac 1{2}-E)^\frac 3{4}(n+\frac 1{2}-E)^\frac 3{4}}.
\end{equation}
The scalar products inside the double series can be estimated using Wang's results on integrals of products of eigenfunctions of the harmonic oscillator \cite{Wa}. While the scalar product clearly vanishes if $m+n=2s+1$, when both indices are either even or odd we get:
\begin{eqnarray}
\left\langle \phi_{2m} \phi_0, \phi_0 \phi_{2n} \right\rangle^2 \!\!&\!\!=\!\!&\!\! \frac 1{2\pi} \left[ \frac { [2(m+n)]!}{(m+n)!} \right]^2 \frac 1{2^{4(m+n)}(2m)!(2n)!}  \nonumber\\
\!\!&\!\! \leq \!\!&\!\!  \frac 1{2\pi} \left[ \frac { [2(m+n)]!}{2^{2(m+n)}[(m+n)!]^2} \right]^2=\frac{\phi_{2(m+n)}^4(0)}{2}, 
\label{8} \\ [1ex]
\left\langle\phi_{2m+1} \phi_0, \phi_0\phi_{2n+1} \right\rangle^2 \!\!&\!\! =\!\!&\!\!  \frac 1{2\pi} \left[ \frac { [2(m+n+1)]!}{(m+n+1)!} \right]^2 \frac 1{2^{4(m+n+1)}(2m+1)!(2n+1)!}  \nonumber\\
\!\!&\!\! \leq \!\!&\!\!  \frac 1{2\pi} \left[ \frac { [2(m+n+1)]!}{2^{2(m+n+1)}[(m+n+1)!]^2} \right]^2=\frac{\phi_{2(m+n+1)}^4(0)}{2},
\label{9}
\end{eqnarray}
the final equalities in \eqref{8} and \eqref{9} resulting from \cite{FI94} and \cite{MS16}. Therefore, the r.h.s. of \eqref{7} is bounded from above by:
\begin{equation}\label{10}
\text{tr}( \tilde{B}_E^2)\leq \frac {\pi^\frac 3{2}}{8}\left[\sum _{m,n=0}^{\infty} \frac {\phi _{2(m+n)}^4(0)}{(2m+\frac 1{2}-E)^\frac 3{4}(2n+\frac 1{2}-E)^\frac 3{4}}+ \sum_{m,n=0}^{\infty}\frac {\phi _{2(m+n+1)}^4(0)}{(2m+\frac 3{2}-E)^\frac 3{4}(2n+\frac 3{2}-E)^\frac 3{4}} \right].
\end{equation} 

As can be gathered from \cite{MS16} using Stirling's formula,
\begin{eqnarray}\label{11}
\phi _{2n}^4(0)\leq\frac 1{\pi^2 n}, && n\geq 1, \nonumber \\
\phi _{2(m+n)}^4(0)\leq\frac 1{\pi^2 (m+n)}, && m,n\geq 1, \nonumber\\
\phi _{2(m+n+1)}^4(0)\leq\frac 1{\pi^2 (m+n+1)}, && m,n\geq 0,
\end{eqnarray} 
which implies that \eqref{10} is bounded by
\begin{eqnarray}
\text{tr}( \tilde{B}_E^2) \!\!&\!\! \leq \!\!&\!\! {\frac 1{8 \pi^\frac 1{2}} \left[\frac 1{(\frac 1{2}-E)^\frac 3{2}}+ \frac 2{(\frac 1{2}-E)^\frac 3{4}}\sum _{n=1}^{\infty }\frac 1{n(2n+\frac 1{2}-E)^\frac 3{4}}\right]    } \nonumber\\
\!\!&\!\!  \!\!&\!\! +
\frac 1{8\pi^\frac 1{2}}\sum _{m=1}^{\infty }\sum _{n=1}^{\infty }\frac 1{m^{\frac 1{2}}(2m+\frac 1{2}-E)^\frac 3{4}{n^{\frac 1{2}}(2n+\frac 1{2}-E)^\frac 3{4}}}  \nonumber\\
\!\!&\!\!  \!\!&\!\! {+
\frac 1{8\pi^\frac 1{2}}\sum _{m=0}^{\infty }\sum _{n=0}^{\infty }\frac 1{(m+\frac 1{2})^{\frac 1{2}}(2m+\frac 3{2}-E)^\frac 3{4}{(n+\frac 1{2})^{\frac 1{2}}(2n+\frac 3{2}-E)^\frac 3{4}}}    } \nonumber\\
\!\!&\!\! \leq \!\!&\!\! 
\frac 1{8 \pi^\frac 1{2}} \left[\frac 1{(\frac 1{2}-E)^\frac 3{2}}+ \frac 2{(\frac 1{2}-E)^\frac 3{4}}\sum _{n=1}^{\infty }\frac 1{n^\frac 1{2}(2n+\frac 1{2}-E)^\frac 3{4}}\right]  \label{12}\\ 
\!\!&\!\!  \!\!&\!\! +
\frac 1{8 \pi^\frac 1{2}}  \left[\sum _{n=1}^{\infty }\frac 1{n^\frac 1{2}(2n+\frac 1{2}-E)^\frac 3{4}}\right]^2 + \frac 1{8\pi^\frac 1{2}}  \left[\sum _{n=0}^{\infty }\frac 1{(n+\frac 1{2})^\frac 1{2}(2n+\frac 3{2}-E)^\frac 3{4}}\right]^2< \infty, \nonumber
\end{eqnarray}
since both series involved in the final expression are clearly absolutely convergent given that the summands are positive sequences decaying like $n^{-\frac 5{4}}$. Hence, the trace of the square of the Birman-Schwinger operator, i.e. its Hilbert-Schmidt norm, is finite for any $E<\frac 1{2}$.

Our result is not surprising at all since the norm could have been bounded by that of the 
Birman-Schwinger operator with the same impurity but with the resolvent of our $H_0$ replaced by that of $-\frac {\Delta}{2}$ in two dimensions, which is known to be finite \cite{AGHH}. However, it provides us with a far more accurate estimate of the norm, which in turn leads to a more precise determination of the spectral lower bound resulting from the use of the Hilbert-Schmidt norm of the Birman-Schwinger operator in the KLMN theorem \cite{RS1}. In fact, the latter bound is what we wish to achieve by further estimating the bottom lines of \eqref{12}.

The series in \eqref{12} can be bounded from above by their respective improper integrals as follows:
\begin{eqnarray}
S_1 \!\!&  \!\!=  \!\! &  \!\! \sum _{n=1}^{\infty }\frac 1{n^\frac 1{2}(2n+\frac 1{2}-E)^\frac 3{4}}=\sum_{n=1}^{\infty }\frac{\sqrt{2}}{(2n)^\frac 1{2}(2n+\frac 1{2}-E)^\frac 3{4}}<  
\int_0^\infty \frac{\sqrt{2}\, dx}{(2x)^\frac 1{2}(2x+\frac 1{2}-E)^\frac 3{4}}  \nonumber\\
\!\!&  \!\!=  \!\! &  \!\!  \frac {3\sqrt{2}}{4}\int_0^\infty \frac {s^\frac 1{2}\, ds}{(s+\frac 1{2}-E)^\frac 7{4}}  \leq 
\frac {3\sqrt{2}}{4}\int_0^\infty \frac{ds}{(s+\frac 1{2}-E)^\frac 5{4}} =\frac {3\sqrt{2}}{(\frac 1{2}-E)^\frac 1{4}}. \label{13} 
\end{eqnarray}
\begin{eqnarray}
S_2 \!\!&  \!\!=  \!\! &  \!\! 
\sum _{n=0}^{\infty }\frac 1{(n+\frac 1{2})^\frac 1{2}(2n+\frac 3{2}-E)^\frac 3{4}}=\frac {\sqrt{2}}{(\frac 1{2}-E)^\frac 3{4}}+\sum _{n=1}^{\infty }\frac{\sqrt{2}}{(2n+1)^\frac 1{2}(2n+\frac 3{2}-E)^\frac 3{4}}  \nonumber\\
\!\!&  \!\! <  \!\! &  \!\! 
\frac {\sqrt{2}}{(\frac 1{2}-E)^\frac 3{4}}+\int_0^\infty \frac{\sqrt{2}\, dx}{(2x+1)^\frac 1{2}(2x+\frac 3{2}-E)^\frac 3{4}} =
\frac {\sqrt{2}}{(\frac 1{2}-E)^\frac 3{4}}-\frac {\sqrt{2}}{(\frac 1{2}-E)^\frac 3{4}}
\nonumber\\
 \!\!&  \!\!  \!\! &  \!\! 
+\frac {3\sqrt{2}}{4}\int_0^\infty \frac {(s+1)^\frac1{2} \, ds}{(s+\frac 3{2}-E)^\frac 7{4}}  \leq 
\frac {3\sqrt{2}}{4}\int_0^\infty \frac{ds}{(s+\frac 3{2}-E)^\frac 5{4}}=\frac {3\sqrt{2}}{(\frac 3{2}-E)^\frac 1{4}}. \label{14}
\end{eqnarray}
Therefore, the bottom lines of \eqref{12} are bounded by:
\begin{eqnarray}
\text{tr}( \tilde{B}_E^2) \!\!&\!\! \leq \!\!&\!\! \frac 1{8 \pi^\frac 1{2}} \left[\frac 1{(\frac 1{2}-E)^\frac 3{2}}+\frac {6\sqrt{2}}{\frac 1{2}-E}+\frac {18}{(\frac 1{2}-E)^\frac 1{2}}\right] +
\frac 1{8 \pi^\frac 1{2}} \frac {18}{(\frac 3{2}-E)^\frac 1{2}}  \nonumber\\
 \!\!&\!\! = \!\!&\!\! 
\frac 1{8 \pi^\frac 1{2} (\frac 1{2}-E)^\frac 1{2}}\left[3\sqrt{2}+ \frac 1{(\frac 1{2}-E)^\frac 1{2}}\right]^2 +
\frac 1{4 \pi^\frac 1{2}} \frac {9}{(\frac 3{2}-E)^\frac 1{2}}. \label{15}
\end{eqnarray}
Hence, our estimate of the Hilbert-Schmidt norm of the Birman-Schwinger operator is:
\begin{eqnarray*}
\text{tr}(\tilde{B}_E^2) \!\!&\!\! = \!\!&\!\!  \left| \left|e^{-\frac {x^2+y^2}{2}} (H_0-E)^{-1} e^{-\frac {x^2+y^2}{2}}  \right|\right|_2^2 \leq 
\frac 1{8 \pi^\frac 1{2}(\frac 1{2}-E)^\frac 1{2}}\left[3\sqrt{2}+ \frac 1{(\frac 1{2}-E)^\frac 1{2}}\right]^2 +
\frac 1{4 \pi^\frac 1{2}} \frac {9}{(\frac 3{2}-E)^\frac 1{2}}. \label{16}
\end{eqnarray*}

As is well known \cite {MFRM,FGNR,KLA,KLA1,FK,F 95,FR 09}, the operator 
$$
(H_0-E)^{-1/2}e^{-(x^2+y^2)} (H_0-E)^{-1/2}
$$
is isospectral to the Birman-Schwinger operator so that their Hilbert-Schmidt norms are identical. Hence, what has been achieved so far can be summarised by means of the following claim.

\begin{theor} 
The integral operators 
$$
(H_0-E)^{-1/2}e^{-(x^2+y^2)} (H_0-E)^{-1/2}\quad \text{and}\quad  e^{-\frac {x^2+y^2}{2}} (H_0-E)^{-1} e^{-\frac {x^2+y^2}{2}}
$$ 
are Hilbert-Schmidt and their Hilbert-Schmidt norms satisfy
\begin{eqnarray}
 \left| \left| (H_0-E)^{-1/2}e^{-(x^2+y^2)} (H_0-E)^{-1/2} \right|\right|_2^2= \left| \left|e^{-\frac {x^2+y^2}{2}} (H_0-E)^{-1} e^{-\frac {x^2+y^2}{2}}  \right|\right|_2^2  \nonumber 
 \\
\quad \leq \frac 1{8 \pi^\frac 1{2} (\frac 1{2}-E)^\frac 1{2}}\left[3\sqrt{2}+ \frac 1{(\frac 1{2}-E)^\frac 1{2}}\right]^2 +
\frac 1{4 \pi^\frac 1{2}} \frac {9}{(\frac 3{2}-E)^\frac 1{2}}. \label{17}
\end{eqnarray}
\end{theor}

As an immediate consequence of the above theorem we get:
\begin{coro} The Hamiltonian 
$$
H_{\lambda}=H_0-\lambda e^{-(x^2+y^2)},
$$
defined in the sense of quadratic forms, is self-adjoint and bounded from below by $E(\lambda)$, the solution of the equation: 
\begin{equation}\label{18}
\frac 1{2 (\frac 1{2}-E)^\frac 1{2}}\left[3\sqrt{2}+ \frac 1{(\frac 1{2}-E)^\frac 1{2}}\right]^2 +
\frac {9}{(\frac 3{2}-E)^\frac 1{2}}=\frac {4 \pi^\frac 1{2}}{\lambda^2} .
\end{equation} 
\end{coro}

\noindent
{\bf Proof.} For any $E<0$ and $\psi\in Q(H_0)=D(H_0^\frac 1{2})$ (the form domain of $H_0$):
\begin{eqnarray}
\!\!&\!\! \!\!&\!\! \lambda \left\langle \psi, e^{-(x^2+y^2)}\psi \right\rangle  = 
\lambda \left\langle (H_0-E)^{1/2}\psi, \left[(H_0-E)^{-1/2}e^{-(x^2+y^2)}(H_0-E)^{-1/2} \right](H_0-E)^{1/2}\psi \right\rangle  \nonumber\\
 \!\!&\!\!  \!\!& \quad \leq
\lambda \left| \left| (H_0-E)^{-1/2}e^{-(x^2+y^2)} (H_0-E)^{-1/2} \right|\right|_2 \left| \left|(H_0-E)^{1/2}\psi  \right|\right|_2^2   \nonumber\\
 \!\!&\!\! \!\!& \quad\leq
\lambda \left[\frac 1{8 \pi^\frac 1{2} (\frac 1{2}-E)^\frac 1{2}}\left[3\sqrt{2}+ \frac 1{(\frac 1{2}-E)^\frac 1{2}}\right]^2 +
\frac 1{4 \pi^\frac 1{2}} \frac {9}{(\frac 3{2}-E)^\frac 1{2}}\right]^\frac 1{2} \left[\left(\psi, H_0\psi \right)-E \left| \left|\psi  \right|\right|_2^2 \right].
\label{19}
\end{eqnarray}
By taking $E$ sufficiently negative, the first factor in the bottom line of \eqref{19} can be made arbitrarily small, which ensures that the Gaussian perturbation is infinitesimally small with respect to $H_0$ in the sense of quadratic forms. Hence, we need only invoke the KLMN theorem (see \cite {RS2}) to infer that  $H_\lambda$ is self-adjoint and bounded from below by the quantity

\begin{equation}\label{lowbo}
\frac{\lambda}{2 \pi^\frac 1{4}} \left[\frac 1{2 (\frac 1{2}-E)^\frac 1{2}}\left[3\sqrt{2}+ \frac 1{(\frac 1{2}-E)^\frac 1{2}}\right]^2 +
 \frac {9}{(\frac 3{2}-E)^\frac 1{2}}\right]^\frac 1{2}E, E<0,
\end{equation} 

\noindent so that the supremum of such lower bounds is attained for that particular value of $E$ solving \eqref{18}.

In the following subsections we first consider a Hamiltonian with a point interaction all along the $x$-direction in place of the Gaussian potential  and then we investigate in detail the solution of \eqref{18}, that is to say the lower bound of the  spectrum of $H_\lambda$.

\subsection{Hamiltonian with a point interaction along the $x$-direction}

\begin{figure}[hbtp]
\begin{center}
\includegraphics[width=0.7\textwidth]{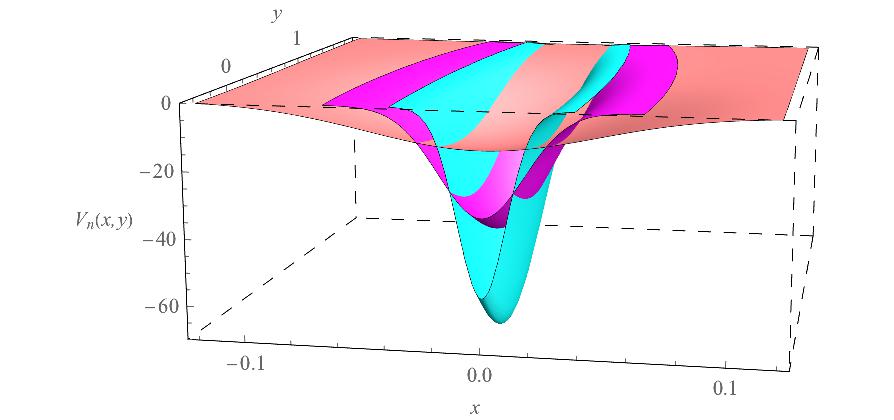}
\caption{\small 
Approximating potentials $ V_n(x,y)=n  e^{-(n^2x^2+y^2)}$ for $n=15$ (pink), $30$ (magenta), and $70$ (cyan).}
\label{fig1}
\end{center}
\end{figure}

Let us consider now the Hamiltonian 
$$
H^\delta_\lambda=H_0-\lambda \sqrt{\pi} \delta(x) e^{-y^2},
$$
that is to say the energy operator given by the same $H_0$ as before but with the interaction term having a point interaction in place of the Gaussian along the $x$-direction. Our goal is to prove that such an operator is self-adjoint and that it is the limit in the norm resolvent sense of the sequence $H_0-\lambda V_n(x,y)$  as $n \rightarrow \infty$, with $V_n(x,y)=n e^{-(n^2x^2+y^2)}$. As is to be expected, our approximating sequence is quite different from the one used in \cite {AGHH} to get the Laplacian perturbed by a point interaction in two dimensions. Before stating and proving the main result of this section, we wish to provide the reader with the visualisation of the approximating potentials in Figure~\ref{fig1}.

\begin{coro} 
The Hamiltonian $H^\delta_\lambda=H_0-\lambda \sqrt{\pi} \delta(x) e^{-y^2}$, defined in the sense of quadratic forms, is self-adjoint and is the norm resolvent limit of the sequence of Hamiltonians 
$$
H_{n,\lambda}=H_0-\lambda n  e^{-(n^2x^2+y^2)}.
$$
Furthermore, $H^\delta_\lambda$ is  bounded from below by  $E_{\delta}(\lambda)$, the solution of the equation: 
\begin{equation}\label{20}
\frac 1{(\frac 1{2}-E)^{\frac 1{2}}}\left[\frac 1{(\frac 1{2}-E)^\frac 1{2}}+3\sqrt{2}\right]^2=\frac {4\pi^{\frac 1{2}}}{\lambda^2}.
\end{equation} 
\end{coro}

\bigskip

\noindent
{\bf Proof.} First of all, it is quite straightforward to show that the integral operator 
$$
(H_0-E)^{-1/2}  \sqrt{\pi}\delta(x) e^{-y^2}(H_0-E)^{-1/2},\quad E<0,
$$is Hilbert-Schmidt with the square of the Hilbert-Schmidt norm given by:
\begin{eqnarray}
&&\!\!\!\!\!\!\!\!\!\! \pi\left| \left| (H_0-E)^{-1/2}\delta(x)e^{-y^2} (H_0-E)^{-1/2} \right|\right|_2^2 \nonumber\\
&&= \pi \left| \left|(\delta(x)e^{-y^2})^\frac 1{2} (H_0-E)^{-1}(\delta(x)e^{-y^2})^\frac 1{2} \right|\right|_2^2 \nonumber\\
&&= \pi \int_{-\infty}^{\infty} \int_{-\infty}^{\infty} e^{-y^2} e^{-y'^2}  \left[(H_0-E)^{-1} (0,0,y,y') \right]^2 dydy' \nonumber\\
&&= \pi \int_{-\infty}^{\infty} \int_{-\infty}^{\infty} e^{-y^2} e^{-y'^2}  \left[\sum _{n=0}^{\infty } \frac {e^{- \sqrt{2 \left(2n+\frac 1{2}-E\right)} \left|y'-y'' \right|}}{\sqrt{2 \left(2n+\frac 1{2}-E\right)}} \phi _{2n}^2(0)\right]^2 dydy' \nonumber\\
&&= \frac {\pi}{2} \sum _{m=0}^{\infty }\sum _{n=0}^{\infty }  \phi _{2m}^2(0)  \phi _{2n}^2(0) \int_{-\infty}^{\infty} \int_{-\infty}^{\infty}e^{-y^2} \frac {e^{-\sqrt{2 \left(2m+\frac 1{2}-E\right)} \left|y-y' \right|}e^{-\sqrt{2 \left(2n+\frac 1{2}-E\right)} \left|y-y' \right|}}{\sqrt{ \left(2m+\frac 1{2}-E\right)\left(2n+\frac 1{2}-E\right)}} e^{-y'^2}dydy'. \nonumber
\end{eqnarray}

As the double integral involving the convolution has already been estimated in \eqref{5m}, the latter expression is bounded by:
\begin{equation}\label{22}
\frac {\pi^{3/2}}{4}\sum _{m=0}^{\infty }\sum _{n=0}^{\infty }\frac {\phi _{2m}^2(0)\phi _{2n}^2(0)}{(2m+\frac 1{2}-E)^\frac 3{4}(2n+\frac 1{2}-E)^\frac 3{4}}=\frac {\pi^{3/2}}{4}\left[\sum _{n=0}^{\infty }\frac {\phi _{2n}^2(0)}{(2n+\frac 1{2}-E)^\frac 3{4}}\right]^2,
\end{equation}
which, using \eqref{12},  is bounded by:
\begin{eqnarray}\label{23}
&& \frac 1{4\pi^{1/2}}\left[\frac 1{(\frac 1{2}-E)^\frac 3{4}}+\sum _{n=1}^{\infty }\frac 1{n^{\frac 1{2}}(2n+\frac 1{2}-E)^\frac 3{4}}\right]^2  \leq\frac 1{4\pi^{1/2}}\left[\frac 1{(\frac 1{2}-E)^\frac 3{4}}+\frac {3\sqrt{2}}{(\frac 1{2}-E)^\frac 1{4}}\right]^2 \nonumber\\
&& \qquad\qquad\qquad\qquad =\frac 1{4\pi^{\frac 1{2}}(\frac 1{2}-E)^{\frac 1{2}}}\left[\frac 1{(\frac 1{2}-E)^\frac 1{2}}+3\sqrt{2}\right]^2,
\end{eqnarray}
having taken advantage of \eqref{13}. As the right hand side of \eqref{23} can be made arbitrarily small by taking $E<0$ large in absolute value, the KLMN theorem ensures, as was done previously in the case of $H_{\lambda}$, the self-adjointness of $H^\delta_\lambda$ as well as the existence of the spectral lower bound $E_{\delta}(\lambda)$ given by the solution of \eqref{20}.

As to the convergence of $H_{n,\lambda}$ to $H^\delta_\lambda$, we start by noting that, for any $E<0$,  the operator $(H_0-E)^{-1/2} ne^{-(n^2x^2+y^2)}(H_0-E)^{-1/2}$ converges weakly to $(H_0-E)^{-1/2}  \sqrt{\pi}\delta(x) e^{-y^2}(H_0-E)^{-1/2}$ as $n \rightarrow \infty$. Furthermore,
\begin{eqnarray}\label{24}
\!\!\!\!\!\!\!\!\!\!\!\!\!&\!\!\!\!\!\!\!&\!\!\!\!\!\!\! \left| \left| (H_0-E)^{-1/2}ne^{-(n^2x^2+y^2)} (H_0-E)^{-1/2} \right|\right|_2^2= \left| \left|ne^{-\frac {n^2x^2+y^2}{2}} (H_0-E)^{-1} e^{-\frac {n^2x^2+y^2}{2}}  \right|\right|_2^2
 \nonumber\\ [1ex]
\!\!\!\!\!\!\!\!\!\!\!&\!\!\!\!\!\!\!&\!\!\!\!\!\!\! = \sum_{l,m=0}^{\infty }\left[ \int_{-\infty}^{\infty}\int_{-\infty}^{\infty}   e^{-y^2}\frac {e^{- \left(\sqrt{2 (l+\frac 1{2}-E)}+ \sqrt{2 (m+\frac 1{2}-E)}\right)\left|y-y' \right|}}{2\sqrt{(l+\frac 1{2}-E)(m+\frac 1{2}-E)}}e^{-y'^2}dydy'\right]\left\langle\phi _{l},n e^{-n^2(\cdot)^2}\phi _{m}\right\rangle^2\!\!.
\end{eqnarray}
Since
\begin{equation}\label{25}
\left\langle\phi_{l},n e^{-n^2(\cdot)^2}\!\phi_{m}\right\rangle  =  n \int_{- \infty}^{\infty} \!e^{-n^2x^2} \!\phi _{l}(x) \phi _{m}(x) dx = 
\int_{- \infty}^{\infty} \!e^{-x^2} \! \phi_{l}(x/n) \phi_{m}(x/n) dx \to \sqrt{\pi} \phi_{l}(0) \phi_{m}(0), \nonumber
\end{equation}
as $n \rightarrow \infty$, the right hand side of \eqref{24} converges to
\begin{eqnarray}
\!\!\!\!\!\!\!&\!\!\!&\!\!\!\!  \frac {\pi}{2} \sum _{l=0}^{\infty }\sum _{m=0}^{\infty } \phi _{2l}^2(0) \phi _{2m}^2(0)\left[ \int_{-\infty}^{\infty}\int_{-\infty}^{\infty}   e^{-y^2}\frac {e^{- \left(\sqrt{2 (2l+\frac 1{2}-E)}+ \sqrt{2 (2m+\frac 1{2}-E)}\right)\left|y-y' \right|}}{\sqrt{(2l+\frac 1{2}-E)(2m+\frac 1{2}-E)}}e^{-y'^2}dydy'\right] \nonumber\\ [1ex]
\!\!\!\!\!\!\!&\!\!\!&  \quad = \pi \left| \left|(\delta(x)e^{-y^2})^\frac 1{2} (H_0-E)^{-1}(\delta(x)e^{-y^2})^\frac 1{2} \right|\right|_2^2 = \pi \left| \left| (H_0-E)^{-1/2}\delta(x)e^{-y^2} (H_0-E)^{-1/2} \right|\right|_2^2.\nonumber
\end{eqnarray}
Hence, the Hilbert-Schmidt norm of $(H_0-E)^{-1/2} ne^{-(n^2x^2+y^2)}(H_0-E)^{-1/2}$ converges to the Hilbert-Schmidt norm of $(H_0-E)^{-1/2}  \sqrt{\pi}\delta(x) e^{-y^2}(H_0-E)^{-1/2}$ as $n \rightarrow \infty$. Due to Theorem 2.21 in \cite {Si}, this fact and the previous weak convergence imply that the convergence actually takes place in the Hilbert-Schmidt norm. Then, the norm convergence of these integral operators ensures the norm resolvent convergence of  $H_{n,\lambda}$ to $H^\delta_\lambda$, as guaranteed by Theorem VIII.25 in \cite {RS1}, which completes our proof of Corollary 2.

\subsection{The lower bound of $\sigma(H_0-\lambda e^{-(x^2+y^2)})$}

\begin{figure}[hbtp]
\begin{center}
\includegraphics[width=0.5\textwidth]{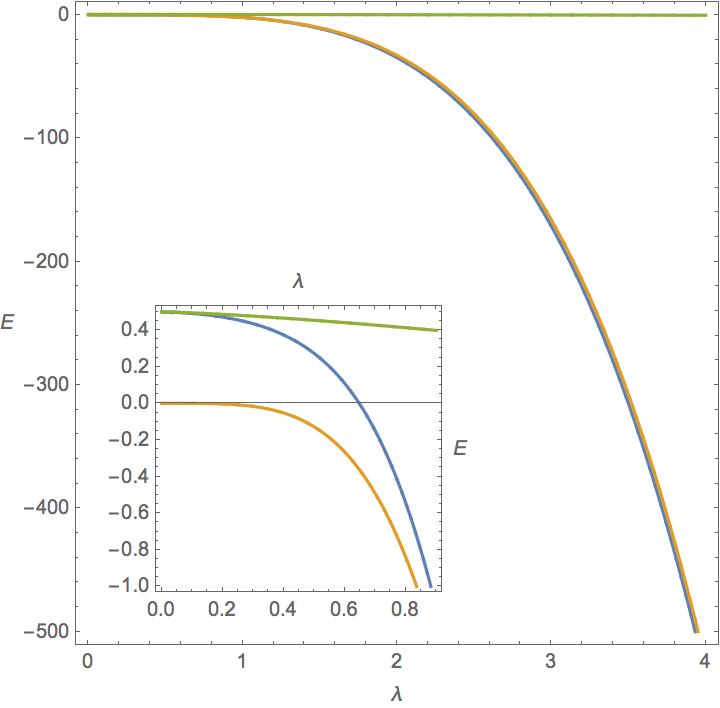}
\caption{\small 
A plot of  the lower bound of the energy $E$ as a function of $\lambda$ resulting from the solution of Equation~\eqref{18} (blue curve), the approximate expression valid for large values of $E$ and $\lambda$ obtained in \eqref{forinfinity} (yellow curve) and the approximation for small values of $\lambda$ as in \eqref{forzero} (green curve). In the inset we have enlarged the region where $\lambda$ and $E$ are small. While the similarity between the solution of \eqref{18} and the funtion \eqref{forinfinity} is quite acceptable for a wide range of the parameters, the solution of \eqref{18} is satisfactorily approximated by \eqref{forzero} only for very small values of $\lambda$.}\label{fig2}
\end{center}
\end{figure}

As anticipated earlier, the lower bound of the  spectrum of $H_\lambda$ in \eqref{1111} is the function  $E(\lambda)$ given implicitly by the equation \eqref{18}. 
From this expression, some approximate results can be easily obtained in two different regimes. For example, it is possible to prove that the asymptotic behaviour of \eqref{18} for large values of both variables  $E(_\lambda),\lambda$ is 
\begin{equation}
E(_\lambda)=- \frac{81}{4\pi^2}\, \lambda^4.
\label{forinfinity}
\end{equation}
On the other hand, for small values of $\lambda$ we can prove that \eqref{18} behaves approximately as follows
\begin{equation}
E(_\lambda)=\frac12- \frac{1}{4\pi^{2/3}}\, \lambda^{4/3}.
\label{forzero}
\end{equation}
A plot of the $\lambda$-dependence ($\lambda$ being the strength of the potential of  $H_{\lambda}$) of the lower bound of the energy $E(_\lambda)$,  resulting from the solution of \eqref{18}, as well as those of the two approximations given by \eqref{forinfinity} and \eqref{forzero}, are given in Figure~\ref{fig2}.

\section{Final remarks}

In this note we have analysed in detail the Birman-Schwinger operator of the two-dimensional Hamiltonian $H_{\lambda}=H_0-\lambda e^{-(x^2+y^2)}$, namely the integral operator $-\lambda e^{-\frac {x^2+y^2}{2}} (H_0-E)^{-1} e^{-\frac {x^2+y^2}{2}}$ where  $H_0=(-\frac 1{2}\frac {d^2}{dx^2}+\frac{x^2}2)-\frac 1{2}\frac {d^2}{dy^2}$. In particular, we have rigorously shown that the operator is Hilbert-Schmidt and have estimated its Hilbert-Schmidt norm. This fact has enabled us to use the KLMN theorem to determine a lower bound for the spectrum of $H_{\lambda}$, that is to say $E(\lambda)$, the implicit function representing the solution of an equation involving the energy parameter and the coupling constant. Furthermore, we have investigated the Hamiltonian $H^\delta_\lambda=H_0-\lambda \sqrt{\pi} \delta(x) e^{-y^2}$, having the Gaussian impurity in the direction subjected to the harmonic confinement replaced by a point impurity.

As anticipated in the introduction, the proof of the resolvent convergence, as the thickness of the layer vanishes, of the Hamiltonian of a three-dimensional parabolic quantum well inside a thin layer to the 2D Hamiltonian \eqref{1111} has been put off as it may deserve a separate paper. 

The results of this article will enable us to study the lowest bound states created by the Gaussian impurity potential of the aforementioned Hamiltonian by means of the modified Fredholm determinant $\text{det}_2 \left[1-\lambda e^{\frac {x^2+y^2}{2}} (H_0-E)^{-1} e^{-\frac {x^2+y^2}{2}} \right]$, the regularised determinant used to handle Hilbert-Schmidt operators. Work in this direction is in progress.

\section*{Acknowledgements}
This work was partially supported by the Spanish MINECO (MTM2014-57129-C2-1-P),
Junta de Castilla y Le\'on and FEDER projects (BU229P18, VA057U16, and VA137G18).

\end{document}